\documentclass{ws-ijmpb}

\newcommand{\med}[1]{\left\langle #1 \right\rangle}
\usepackage{color}

\begin{document}

\markboth{Authors' Names}
{Instructions for Typing Manuscripts (Paper's Title)}

\catchline{}{}{}{}{}
%
%%%%%%%%%%%%%%%%%%%%%%%%%%%%%%%%%%%%%%%%%%%%%%%%%%%%%%%%%%%%%%%%%%%%

%\title{Quantum correlations at finite temperatures as a detector of quantum phase transitions}

%\author{T. Werlang}
%
%\address{Departamento de Fisica, Universidade Federal de S\~ao Carlos, S\~ao Carlos, SP 13565-905, Brazil}
%
%\author{G. A. P. Ribeiro}
%
%\address{Departamento de Fisica, Universidade Federal de S\~ao Carlos, S\~ao Carlos, SP 13565-905, Brazil}
%
%\author{Gustavo Rigolin\footnote{rigolin@ufscar.br}}
%
%\address{Departamento de Fisica, Universidade Federal de S\~ao Carlos, S\~ao Carlos, SP 13565-905, Brazil}

\title{Interplay between quantum phase transitions and the behavior of quantum correlations at finite temperatures}

\author{T. Werlang, G. A. P. Ribeiro, and Gustavo Rigolin\footnote{rigolin@ufscar.br}}

\address{Departamento de F\'{\i}sica, Universidade Federal de S\~ao Carlos, S\~ao Carlos, SP 13565-905, Brazil}

\maketitle

\begin{history}
\received{Day Month Year}
\revised{Day Month Year}
%\accepted{(Day Month Year)}
%\comby{(xxxxxxxxxx)}
\end{history}

\begin{abstract}
We review the main results and ideas showing that quantum correlations at finite temperatures ($T$), 
in particular quantum discord, are useful tools 
in characterizing quantum phase transitions that only occur, in principle, 
at the unattainable absolute zero temperature.  
We first review some interesting results about the behavior of
thermal quantum discord for small spin-1/2 chains and show that they already give us important hints
of the infinite chain behavior. We then study in detail and in the thermodynamic limit (infinite chains)
the thermal quantum correlations for
the XXZ and XY models, where one can clearly appreciate 
that
%
%why 
the behavior of
thermal quantum discord at finite $T$ is a useful tool to spotlight the critical point of
a quantum phase transition.
\end{abstract}


\section{Introduction}

A deeper understanding of the low temperature macroscopic phases of a many-body system 
can only be achieved with the aid of a quantum theory. %%quantum mechanics/quantum field theory
%%A deeper understanding of the many macroscopic phases that a many-body system 
%%can have at low temperatures can only be achieved with the aid of quantum mechanics.
%Quantum mechanics plays an important role in the description of the phases of 
%a many-body system at low temperatures. 
Of particular interest is the modeling and description of how the system goes
from one phase to another, a process known as ``phase transition''.
In ordinary phase transitions, 
which occur at finite temperatures, the %macroscopic order is destroyed by 
phase change is driven by thermal fluctuations. For example, if we heat a magnet 
we arrive at a temperature (Curie temperature) above which it loses its magnetism. In other
words, the magnet initially in the ferromagnetic phase, where all spins are aligned,
changes to the paramagnetic phase, where the magnetic moments are in a disordered state.

However, a phase transition can also occur at or near absolute zero temperature ($T=0$), where 
thermal fluctuations are negligible. This process is called 
a quantum phase transition (QPT)\cite{sac99} and is attained %obtained 
varying a tuning parameter in the 
Hamiltonian (e.g. an external magnetic field) while keeping the temperature
fixed. When the tuning parameter reaches a particular 
value, the so called critical point (CP), the system's Hamiltonian ground state changes 
drastically, which reflects in an abrupt modification of the macroscopic properties of the system. 
In this scenario, quantum fluctuations, which loosely speaking are governed by the 
Heisenberg uncertainty principle, are responsible 
for the phase transition. %destroying the macroscopic order. 
Although it is impossible to 
achieve $T=0$ due to the third law of thermodynamics, 
the effects of QPTs can be observed at finite temperatures 
whenever the de Broglie wavelength is greater 
than the correlation length of thermal fluctuations\cite{sac99}. 
Important examples of QPTs are 
the paramagnetic-ferromagnetic transition in some metals\cite{row10}, 
the superconductor-insulator transition\cite{dol10}, 
and superfluid-Mott insulator transition\cite{gre02}.

The ground state of a many-body system at $T=0$ near a CP is often described by a %complex 
non-trivial wave function due to the long-range correlations among the system's constituents. 
In Ref.~\refcite{pres00} Preskill argued that quantum entanglement 
could be responsible for these correlations and therefore the methods developed by quantum information 
theory (QIT) could be useful in studying the critical behavior of many-body systems. Besides, 
new protocols for quantum computation and communication could be formulated based on such systems. 
Along these lines many theoretical tools (in particular entanglement quantifiers)
originally developed to tackle QIT problems
were employed to determine the CPs of QPTs at $T=0$ \cite{Nie99}. 
Later L.-A. Wu et al.\cite{lidar} proved a general connection between non-analyticities in 
bipartite entanglement measures and QPTs while in Ref.~\refcite{oliveira} 
this connection was extended to multipartite entanglement.

Recently another quantity, namely, quantum discord (QD), has attracted the attention of the 
quantum information community. QD goes beyond the concept of entanglement and captures
in a certain sense the ``quantumness'' of the correlations between two parts of a system. 
It was built by noting the fact that two classically 
equivalent versions of the mutual information are inequivalent in the quantum domain \cite{zurek,vedral}. 
Using QD one can show that quantum entanglement does not describe all quantum correlations existing 
in a correlated quantum state. In other words, it is possible to create quantum correlations 
other than entanglement between two quantum systems via local operations and classical communication (LOCC). 

The first study showing a possible connection between QD and QPT was done by
R. Dillenschneider\cite{Dil08}  in the context of spin chains at $T=0$, 
where a CP associated to a QPT was well characterized by QD. 
Further studies subsequently have confirmed the usefulness of QD in 
describing other types of QPTs at zero temperature\cite{disT0}. 
Moreover, it is important to note 
that such quantum informational approaches to spotlight CPs of QPTs do not
require the knowledge of an order parameter (a macroscopic quantity 
that changes abruptly during the QPT); only the extremal values or 
the behavior of the derivatives of either the entanglement or QD is sufficient.

The previous theoretical studies, however, were restricted to the zero temperature 
regime, which is 
experimentally unattainable
% an unphysical assumption
; the third law of thermodynamics
dictates that it is impossible to drive a system to $T=0$ by a finite amount of 
thermodynamic operations. 
Due to this limitation it is therefore not straightforward to directly compare 
those theoretical results with experimental data obtained at finite $T$. 
In order to overcome this problem %the first step is 
one has to study the behavior of 
quantum correlations in a system at thermal equilibrium, 
which is described by the canonical ensemble 
$\rho_T=e^{-H/kT}/Z$, with $H$ being the system's Hamiltonian, $k$ the Boltzmann's 
constant and $Z=\mbox{Tr}\left(e^{-H/kT}\right)$ the partition function. 
{\it Thermal entanglement}, i.e. the entanglement computed for states described 
by $\rho_T$, was first studied by M. C. Arnesen {\it et al.}\cite{Ved01} for 
a finite unidimensional Heisenberg chain. Other interesting works followed this one,
where thermal entanglement has been considered for other Hamiltonians, both 
for finite\cite{Kam02,Rig03} and infinite chains\cite{ltte}. See Ref. \refcite{reviews}
for extensive reviews on entanglement and QPT. 
However, the focus of the aforementioned works was not the study of the ability of 
thermal entanglement to point out CPs when $T>0$. 
In Ref.~\refcite{Wer10}, two of us introduced the analogous to thermal entanglement, 
namely, {\it thermal quantum discord} (TQD), and we studied the behavior of this quantity 
in a system consisting of two spins described by the $XYZ$ model in the presence of an
external magnetic field. In this work it was observed for the first time that 
TQD could be able to signal a QPT at \textit{finite} $T$.

In order to fully explore the previous possibility, highlighted by solving 
a simple two-body problem\cite{Wer10}, we tackled   
the XXZ Hamiltonian in the thermodynamic 
limit for several values of $T>0$\cite{werPRL}. Now, working with infinite chains,
we were able to show for the
first time that TQD keeps its ability to 
detect the CPs associated to the QPTs for the XXZ model even if $T\neq 0$, while 
entanglement and other thermodynamic quantities are not as good as TQD. Also,
we showed that these quantities when contrasted to TQD lose for increasing $T$ their 
CP-detection property faster than TQD. Later\cite{werPRA}
we generalized those results considering $(i)$ the XXZ Hamiltonian and
$(ii)$ the XY Hamiltonian both in the presence 
of an external magnetic field. For these two models it was shown that 
among the usual quantities employed to detect CPs, TQD was the best suited to
properly estimate them when $T>0$.

Our goal in this paper is to present a short but self-contained review of 
our aforementioned results
%the aforementioned works of ours 
about TQD and its application as a CP detector of QPTs.
%, that in principle occur at $T=0$, studying the system at finite values of $T$.
To this end we structure this paper as follows.
In Sec. 2 we present a brief review about quantum correlations 
where QD and the entanglement of formation take on a prominent role. 
In Sec. 3 we review the behavior of TQD in the context of simple two-qubit models. 
We then move on to the analysis of the ability of quantum correlations,
in particular TQD, to spotlight the CPs of QPTs when the system is 
at $T>0$ and in the thermodynamic limit. In this context we study the
XXZ and XY models with and without an external magnetic field. 
Finally, we conclude and discuss future directions in Sec. 4.

\section{Quantum Correlations}

The superposition principle of quantum mechanics is directly related to
the existence of (quantum) correlations that are not seen 
in classical objects. This principle together with the tensorial nature of
combining different quantum systems (Hilbert spaces) lead to entanglement,
which implies intriguing correlations among
the many constituents of a composite system that puzzle our classical minds. 
It is worth mentioning that this tensorial nature from which a composite quantum
system is described in terms of its parts is not a truism. Indeed, the principle of
superposition is also present in classical physics, for example in the classical theory 
of electromagnetism. However, in classical physics this tensorial nature for combining systems is missing, 
which helps in understanding why ``weird'' quantum effects such as non-locality\cite{EPR} are not
seen in a classical world.

During many decades after the birth of quantum mechanics 
quantum correlations were thought to be necessarily linked to non-locality, 
or more quantitatively, to the violation of a Bell-like inequality\cite{Bel64}.
The non-violation of a Bell-like inequality implies that 
the correlations among the parts of a composite system
can be described by a local realistic theory. 
This fact led many to call a state 
not violating any Bell-like inequality a classical state.

This situation changed by the seminal work of R. F. Werner\cite{Wer89}, who showed 
that there are mixed entangled states that do not violate any Bell-like inequality. 
Therefore, according to the Bell/non-locality paradigm these states should be considered 
examples of classical states although possessing entanglement. 
This state of affairs was unsatisfactory and the notion of classical states was
expanded. A classical (non-entangled) state was then defined as any state 
that can be created only by local operations on the subsystems and classical communication
among its many parts (LOCC)\cite{nielsen,Hor09}. For a  bipartite system described by the density 
operator $\rho_{AB}$, the states created via LOCC (separable states) can 
be generally written as
\begin{eqnarray}\label{sstate}
\rho_{AB}=\sum_jp_j\rho_j^A\otimes\rho_j^B,
\end{eqnarray}
where $p_j\geq 0$, $\sum_jp_j=1$, and $\rho_j^{A,B}$ are legitimate density matrices. 
If a quantum state cannot be written as (\ref{sstate})
then it is an entangled state. 
%Using this definition Horodecki et al.\cite{Hor09} 
%showed that the non-classical states are exactly the entangled quantum states. 

At this point one may wonder if this is a definitive characterization of a classical state.
Or one may ask: Isn't there any ``quantumness'' in the correlations for some sort of 
separable (non-entangled) quantum states? Can we go beyond the entanglement paradigm?
As observed in refs.~\refcite{zurek,vedral} there exist some states written as (\ref{sstate})
that possess non-classical features. This fact led the authors of 
refs.~\refcite{zurek,vedral,ved10b} to push further our definition of classical states. 
Now, instead of eq. (\ref{sstate}), we call a bipartite 
state classical if it can be written as
\begin{eqnarray}
\rho_{AB}=\sum_{jk}p_{jk}|j \rangle_A \langle j|\otimes |k\rangle_B\langle k|,
\label{two}
\end{eqnarray}
where $|j\rangle_A$ and $|k\rangle_B$ span two sets of orthonormal states. 
States described by (\ref{two}) are a subset of those described by (\ref{sstate})
and they are built via mixtures of locally distinguishable states\cite{ved10b}.
Intuitively, classical states are those where the superposition principle does
not manifest itself either on the level of different Hilbert spaces (zero entanglement) 
or on the level of single Hilbert spaces (no Schr\"odinger cat states leading to a mixture
of non-orthogonal states). Such states have null QD.  

Let us be more quantitative and define QD for a bipartite quantum 
state \cite{zurek,vedral} divided into parts $A$ and $B$.
In the paradigm of classical information theory\cite{nielsen} 
the total correlation between $A$ and $B$ is quantified by the mutual information (MI),
\begin{eqnarray}\label{mi1}
\mathcal{I}_1(A:B)= \mathcal{H}(A)+\mathcal{H}(B)-\mathcal{H}(A,B),
\end{eqnarray}
where $\mathcal{H}(X)=-\sum_xp_x\log_2p_x$ is the Shannon entropy with $p_x$ 
the probability distribution of the random variable $X$. The conditional 
probability for classical variables $p_{a|b}$ is defined by the Baye's rule, 
that is, $p_{a|b}=p_{a,b}/p_{b}$, with $p_{a,b}$ denoting the joint probability distribution
of variables $a$ and $b$. Using the conditional probability one  can show that
$\mathcal{H}(A,B) = \mathcal{H}(A|B) + H(B)$, where $\mathcal{H}(A|B)=-\sum_{a,b}p_{a,b}\log_2(p_{a|b})$.
This last result allows one to write MI, 
eq.~(\ref{mi1}), as     
\begin{eqnarray}\label{mi2}
\mathcal{I}_2(A:B)= \mathcal{H}(A)-\mathcal{H}(A|B).
\end{eqnarray}
Note that $\mathcal{H}(A|B)\geq0$ 
is the conditional entropy, which quantifies 
how much uncertainty is left on average about $A$ when one knows $B$. 
The quantum version of eq. (\ref{mi1}), denoted by $\mathcal{I}^q_1(A:B)$, 
can be obtained replacing the Shannon entropy by the von-Neumann entropy 
$\mathcal{S}(X)=\mathcal{S}(\rho_X)=-\mbox{Tr}\left(\rho_X\log_2\rho_X\right)$, 
where $\rho_X$ is the density operator of the system $X=A,B$. On the other 
hand, a quantum version of eq. (\ref{mi2}) is not so straightforward because 
the Bayes' rule is not always valid for quantum systems\cite{brule}. 
For instance, this rule is violated for a pure entangled state. Indeed, 
one can show that if we naively extend the usual quantum conditional entropy
to quantum systems as $\mathcal{S}(A|B)\equiv\mathcal{S}(A,B)-\mathcal{S}(B)$,
it becomes negative for pure entangled states. QD is built in a way to
circumvent this limitation.

In order to build a meaningful quantum version of the conditional entropy 
$\mathcal{H}(A|B)$ it is necessary to take into account the fact that knowledge about
system B is related to measurements on $B$. 
And now, differently from the classical case, a measurement in the quantum domain 
can be performed in many non-equivalent ways (different set of projectors, for instance). 
If a general quantum measurement, i.e. a POVM (positive operator valued measure)\cite{nielsen},
$\{M_b\}$ is performed 
in the quantum state $\rho_{AB}$, then after the measurement the state is
described by $\sum_bM_b\rho_{AB}M_b^\dagger$. 
The probability of the outcome $b$ of $B$ is $p_b=\mbox{Tr}\left[M_b\rho_{AB}M_b^\dagger\right]$ and 
the conditional state of $A$ in this case is 
$\rho_{A|b}=\left(M_b\rho_{AB}M_b^\dagger\right)/p_b$. Thus, the conditional 
entropy with respect to the POVM $\{M_b\}$ is $\mathcal{S}(A|\{M_b\})\equiv\sum_bp_b\mathcal{S}(\rho_{A|b})$. 
Therefore, in order to quantify the uncertainty left on $A$ after a measurement on $B$ 
one has to minimize over all POVMs. This leads to the following definition 
of the quantum conditional entropy\cite{zurek,vedral}:
\begin{eqnarray}\label{qce}
\mathcal{S}_q(A|B)\equiv\min_{\left\{M_b\right\}}\mathcal{S}(A|\{M_b\}).
\label{cond}
\end{eqnarray}  
 
At this point, the quantum version of the mutual information (\ref{mi2}), 
denoted by $\mathcal{I}^q_2(A:B)$, is obtained replacing the classical 
conditional entropy $\mathcal{H}(A|B)$ by its quantum analog $\mathcal{S}_q(A|B)$, 
which does not assume negative values.  The {\it quantum discord} is 
defined as the difference between these two versions of the quantum mutual 
information\cite{zurek,vedral}:
\begin{eqnarray}\label{qd}
D(A|B)&\equiv&\mathcal{I}^q_1(A:B)-\mathcal{I}^q_2(A:B)\nonumber\\
&=&\mathcal{S}_q(A|B)-\mathcal{S}(A|B).
\end{eqnarray} 
Note that QD is not necessarily a symmetric quantity, because the 
conditional entropy (\ref{qce}) depends on the system in which the 
measurement is performed. However, for the density operators studied 
in this paper QD is always symmetric. Furthermore, while for mixed 
states there are states with null entanglement and QD $>0$, for pure 
states QD is essentially equivalent to entanglement. In other words,
only for mixed 
states there might be quantum correlations other than entanglement. 
As demonstrated recently\cite{ved10}, a quantum state $\rho_{AB}$ has $D(A|B)=0$ if, 
and only if, it can be written as 
$\rho_{AB}=\sum_jp_j\rho_j^A\otimes\left|\psi_j^B\right\rangle\left\langle \psi_j^B\right|$, 
with $\sum_jp_j=1$ and $\{\left|\psi_j^B\right\rangle\}$ a set of orthogonal states. 
This result shows the importance of the superposition principle 
to explain the origin of the quantum correlations. It is due to 
this principle that one can generate a set $\{\left|\psi_j^B\right\rangle\}$ 
of non-orthogonal states leading to states with nonzero QD.

For arbitrary $N\times M$-dimensional bipartite states the computation of QD involves a complicated 
minimization procedure whose origin can be traced back to the evaluation of 
the conditional entropy $\mathcal{S}_q(A|B)$, eq. (\ref{cond}). 
In general one must then rely on numerical procedures to get QD 
and it is not even known whether a general efficient algorithm exists.   
For two-qubit systems, however, the minimization over generalized measurements 
can be replaced by the minimization over projective measurements 
(von Neumann measurements)\cite{minDIS}. In this case the minimization 
procedure can be efficiently implemented numerically\cite{Dnum} for arbitrary 
two-qubit states and some analytical results can be achieved for a restricted 
class of states\cite{Danalitico}. In this work we will be dealing with density 
matrices $\rho$ in the X-form, that is, $\rho_{12}=\rho_{13}= \rho_{24}=\rho_{34}=0$. 
Moreover, in our models $\rho_{22}=\rho_{33}$ and all matrix elements are real, making the numerical
evaluation of QD simple and fast. 

To close this section, we introduce the measure of entanglement 
used in this paper, the {\it Entanglement of Formation} (EoF)\cite{Woo98}. 
EoF quantifies, at least for pure states, 
how many singlets are needed per copy of $\rho_{AB}$ 
to prepare many copies of $\rho_{AB}$ using only LOCC.  For an $X$-form 
density matrix we have
\begin{eqnarray}\label{eof}
EoF(\rho_{AB})&=&-g\log_2g-(1-g)\log_2(1-g),
\end{eqnarray}
with $g=(1+\sqrt{1-C^2})/2$ and the concurrence\cite{Woo98} 
given by $C=2\max\left\{0,\Lambda_1,\Lambda_2 \right\}$,
where $\Lambda_1=|\rho_{14}|-\sqrt{\rho_{22}\rho_{33}}$ and 
$\Lambda_2=|\rho_{23}|-\sqrt{\rho_{11}\rho_{44}}$.

\section{Results and Discussions}

\subsection{Two interacting spins}

In this section we consider a two 
spin system described by the XYZ model with an external magnetic 
field acting on both spins.\cite{Wer10} 
The Hamiltonian of this model is
\begin{eqnarray}\label{hxyz}
H_{XYZ}=\frac{J_x}{4}\sigma^x_1\sigma^x_{2}+\frac{J_y}{4}\sigma^y_1\sigma^y_{2}+\frac{J_z}{4}\sigma^z_1\sigma^z_{2} 
+ \frac{B}{2}\left(\sigma^z_1+\sigma^z_2\right),
\end{eqnarray} 
where $\sigma_j^\alpha$ ($\alpha=x,y,z$) are the 
usual Pauli matrices acting on the $j$-th site and 
we have assumed $\hbar=1$. As mentioned above, the 
density matrix describing a system in equilibrium with 
a thermal reservoir at temperature $T$ is $\rho_T=e^{-H/kT}/Z$, 
where $Z=\mbox{Tr}\left(e^{-H/kT}\right)$ is the partition function.  
Therefore, the thermal state for the Hamiltonian (\ref{hxyz}) assumes the following form
\begin{equation}
\rho  =  \frac{1}{Z} \left(
\begin{array}{cccc}
A_{11} & 0 & 0 & A_{12}\\
0 & B_{11}  & B_{12} & 0 \\
0 & B_{12} & B_{11} & 0 \\
A_{12} & 0 & 0 & A_{22} \\
\end{array}
\right), \label{rho2}
\end{equation}
with $A_{11}$  $=$ $\mathrm{e}^{-\alpha}$ $(\cosh(\beta)$ $-$ $4B$ $\sinh(\beta)/\eta)$, 
$A_{12}$ $=$  $-$ $\Delta$ $\mathrm{e}^{-\alpha}$ $\sinh(\beta)/\eta$, 
$A_{22}$ $=$ $\mathrm{e}^{-\alpha}$ $(\cosh(\beta)$ $+$ $4$ $B$ $\sinh(\beta)/\eta)$, 
$B_{11}$ $=$ $\mathrm{e}^{\alpha}$
$\cosh(\gamma)$, $B_{12}$ $=$ $-$ $\mathrm{e}^{\alpha}$ $\sinh(\gamma)$, 
and $Z = 2\left( \exp{(-\alpha)}\cosh(\beta) + \exp{(\alpha)}\cosh(\gamma) \right)$,
where $\Delta = J_{x} - J_{y}$, $\Sigma = J_{x} + J_{y}$,
$\eta = \sqrt{\Delta^{2} + 16B^2}$, $\alpha = J_{z}/(4kT)$, 
$\beta = \eta/(4kT)$, and $\gamma = \Sigma/(4kT)$.

\begin{figure}%[!tb]
\centerline{\psfig{file=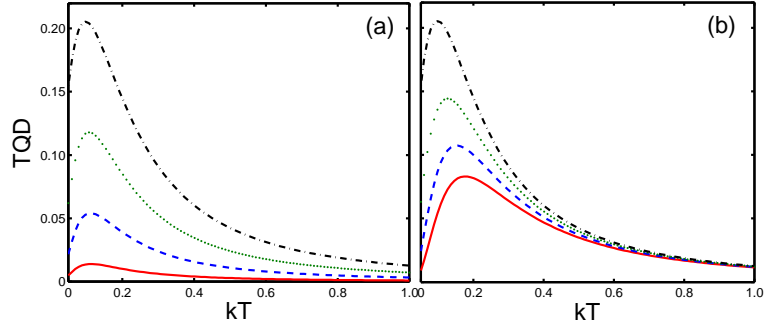,width=4.00in}}
%\vspace*{8pt}
\caption{TQD for a two spin system as a function of the absolute temperature $kT$ for 
the $XXZ$ model with $B=0$. (a) Here $J_z=-0.5$ and $J$ $=$ $0.1$ (solid line), 
$0.2$ (dashed line), $0.3$ (dotted line), $0.4$ (dash-dotted line); 
(b) Now we fix $J=0.4$ and $J_z=-0.8$ (solid line), $-0.7$ (dashed line), 
$-0.6$ (dotted line), $-0.5$ (dash-dotted line). Here and in the
following graphics all quantities are dimensionless.}
\label{fig1}
\end{figure}

The first important result appears in the absence of an external 
field. As shown in Ref.~\refcite{Rig03}, when $B=0$, the entanglement 
does not increase with increasing temperature. On the other 
hand, as can be seen in Fig.~\ref{fig1} (panels $a$ and $b$) for the 
XXZ model $\left(J_x=J_y=J \quad\mbox{and}\quad J_z\neq0 \right)$, 
TQD begins with a non-null value at $T=0$ and increases as $T$ increases
before decreasing with $T$, 
while EoF is always zero\cite{Rig03}.
Note that such effect is observed 
for different configurations of coupling constants. 

\begin{figure}%[!bt]
\centerline{\psfig{file=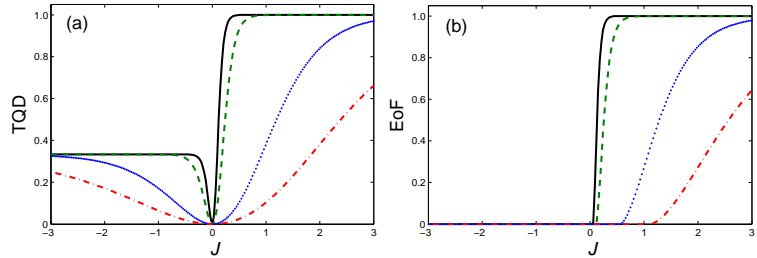,width=4.00in}}
%\vspace*{8pt}
\caption{TQD (a) and EoF (b) for a two spin system as functions of the coupling $J$ 
for $kT=0.05$ (solid line), $0.1$ (dashed line), $0.5$ (dotted line), 
$1.0$ (dash-dotted line). Both plots for the XXX model with $B=0$.}
\label{fig2}
\end{figure}

The possibility of 
TQD to point out a QPT even when the system is at finite temperatures emerged from 
our study about the XXX model for two spins\cite{Wer10}. The XXX model 
is obtained from Hamiltonian (\ref{hxyz}) making $J_x=J_y=J_z=J$. 
When $J\rightarrow\infty$ the density operator (\ref{rho2}) will be 
the Bell state $\rho=\left|\psi\right\rangle\left\langle\psi\right|$, 
with $\left|\psi\right\rangle=\frac{1}{\sqrt{2}}\left(\left|01\right\rangle-\left|10\right\rangle\right)$, 
for any $T$. For the opposite limit $J\rightarrow-\infty$ the density 
operator is the mixed state $\rho=\frac{1}{3}\left(\left|00\right\rangle\left\langle
00\right|+\left|11\right\rangle\left\langle11\right|+\left|\phi\right\rangle\left\langle \phi\right|\right)$ 
with $\left|\phi\right\rangle=\frac{1}{\sqrt{2}}\left(\left|01\right\rangle +\left|10\right\rangle\right)$. 
In this case EoF is zero, while TQD assumes the value $1/3$. 
Furthermore, as shown in Fig.~\ref{fig2}, EoF is zero in the 
ferromagnetic region $(J<0)$ and non-zero in the antiferromagnetic 
region $(J>0)$ only when $T=0$. 
For $T>0$ EoF becomes non-zero 
only for $J>J_c(T)=kT\ln(3)$. On the other hand, TQD is equal 
to zero only at the 
trivial
%
%critical 
point $J=0$, even at finite $T$. 
Although we are considering here only two spins, such result 
suggests that TQD may possibly signal a QPT for $T>0$.

\begin{figure}%[!bt]
\centerline{\psfig{file=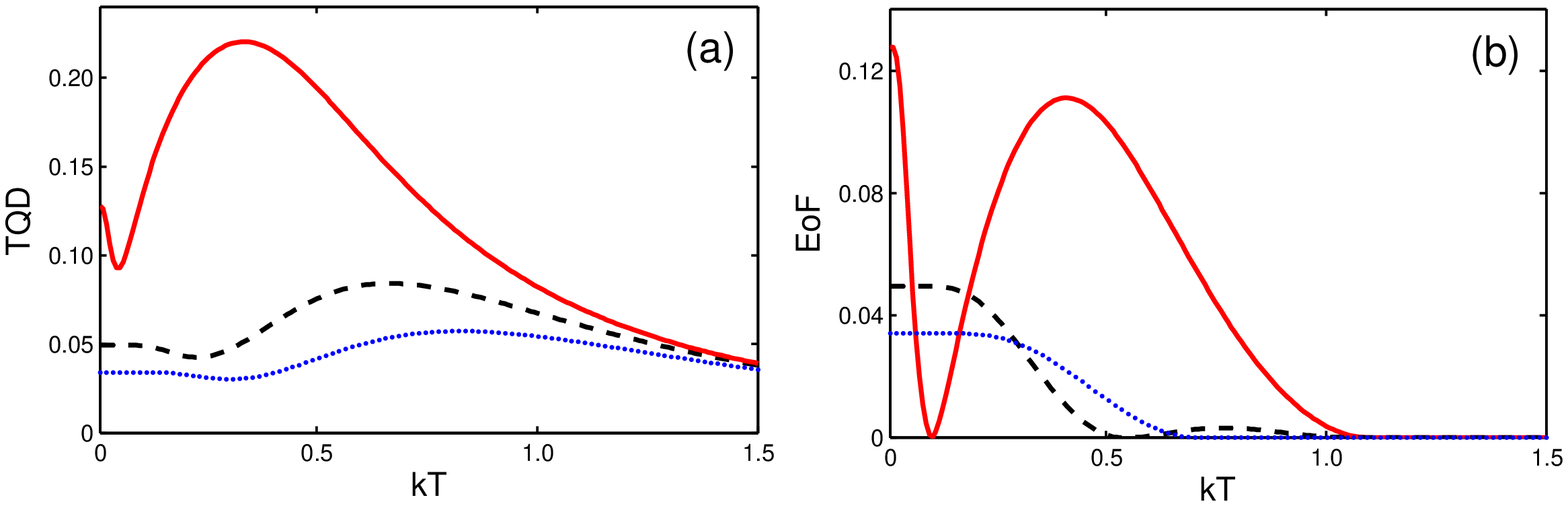,width=4.00in}}
%\vspace*{8pt}
\caption{TQD (a) and EoF (b) for a two spin system as functions of the absolute temperature 
$kT$ for the XY model with transverse magnetic field $B$. Here $J_x=2.6$ 
and $J_y=1.4$. The values for $B$ are $1.1$ (solid line), 
$2.0$ (dashed line), and $2.5$ (dotted line).}
\label{fig3}
\end{figure}

Let us now analyze the case where $B\neq 0$ and focus on 
the XY model in a transverse magnetic field ($J_x$, $J_y\neq 0$, and $J_z=0$). 
As noted in Ref.~\refcite{Rig03} 
EoF shows a sudden death and then a revival (see Fig.~\ref{fig3}, panel b). 
However, TQD does not suddenly disappear as Fig.~\ref{fig3}, panel a, 
depicts. Actually, TQD decreases with $T$ to a non-null value 
and after a critical temperature $T_c$ 
it starts increasing again. This effect is called {\it regrowth}\cite{Wer10}. 
Although the regrowth of EoF with temperature is not observed for 
two spin chains, we showed in Ref.~\refcite{werPRA} that such interesting 
behavior is possible in the thermodynamic limit. Also, if we carefully look at 
Fig.~\ref{fig3} the distinctive aspects of these two types of quantum correlations 
become more evident. For example, comparing panels a and b we see 
regions where TQD increases while EoF decreases.  Finally, in a very interesting
and recent work, X. Rong \textit{et al.}\cite{Ron12} experimentally verified some of the 
predictions here revised and reported in refs.~\refcite{Wer10,werPRL,werPRA}, namely, 
the sudden change of TQD at finite temperatures while 
changing the anisotropy parameter of a two spin XXZ Hamiltonian.

\subsection{XXZ Model}

Turning our attention to infinite chains ($L\rightarrow \infty$), let us start
working with the one-dimensional anisotropic spin-$1/2$ 
XXZ model subjected to a magnetic field in the $z$-direction. 
Its Hamiltonian is
\begin{eqnarray}\label{hxxz}
H_{xxz}&=&J\sum_{j=1}^L\left(\sigma_j^x\sigma_{j+1}^x+\sigma_j^y\sigma_{j+1}^y
+\Delta\sigma_j^z\sigma_{j+1}^z\right) - \frac{h}{2}\sum_{j=1}^L\sigma_j^z,
\end{eqnarray}
where $\Delta$ is the anisotropy parameter, $h$ is the external 
magnetic field, and $J$ is the exchange constant ($J=1$). We have 
assumed periodic boundary conditions $\left(\sigma_{L+1}^\alpha=\sigma_1^\alpha\right)$. 
The nearest neighbor two spin state is obtained by 
tracing all but the first two spins, $\rho_{1,2}=\mbox{Tr}_{L-2}(\rho)$, 
where $\rho=\exp{\left(-\beta H_{xxz}\right)}/Z$. The Hamiltonian (\ref{hxxz}) 
exhibits both translational invariance and $U(1)$ invariance $\left(\left[H_{xxz},
\sum_{j=1}^L\sigma_j^z\right]=0\right)$ leading to the following nearest neighbor 
two spin state
\begin{eqnarray}\label{rho}
\rho_{1,2}  = \frac{1}{4}\left(
\begin{array}{cccc}
\rho_{11} & 0 & 0 & 0\\
0 & \rho_{22}  & \rho_{23} & 0 \\
0 & \rho_{23} & \rho_{22} & 0 \\
0 & 0 & 0 & \rho_{44}  \\
\end{array}
\right),
\end{eqnarray}
where
\begin{eqnarray}\label{rhoe}
\rho_{11} &=& 1+2\left\langle \sigma^z\right\rangle+\med{\sigma_1^z\sigma_2^z},\nonumber\\
\rho_{22} &=& 1-\med{\sigma_1^z\sigma_2^z},\\
\rho_{44} &=& 1-2\left\langle \sigma^z\right\rangle+\med{\sigma_1^z\sigma_2^z},\nonumber\\
\rho_{23} &=& 2\med{\sigma_1^x\sigma_2^x}\nonumber.
\end{eqnarray}

The magnetization and the two-point correlations above are obtained in terms 
of the derivatives of the free-energy\cite{NLIE}, $f=-\frac{1}{\beta} 
\lim_{L\rightarrow \infty} \frac{\ln{Z}}{L}$, 
\begin{eqnarray*}
\med{\sigma^z}&=& -2\partial_h f/J, \\
\med{\sigma_j^{z}\sigma_{j+1}^{z}}&=&\partial_{\Delta}f/J,  \\
\med{\sigma_j^{x}\sigma_{j+1}^{x}}&=&\frac{u-\Delta \partial_{\Delta}f+ 
h \med{\sigma^z}}{2J}, \\ 
\med{\sigma_j^{z}\sigma_{j+1}^{z}}&=&\med{\sigma_j^{x}\sigma_{j+1}^{x}}=
\frac{u + h \med{\sigma^z}}{3J}, ~~ \Delta=1,
\end{eqnarray*}
where $u=\partial_{\beta} (\beta f)$ is the internal energy. We explained 
in details the procedures to determine the free-energy $f$ in Ref.~\refcite{werPRA}. 
It is not a simple task and it involves the application of complicated 
analytical and numerical computations. The CPs of the XXZ 
model depend on the value of the magnetic field $h$\cite{GAUDIN}. 
One of them, called $\Delta_{inf}$, is an infinite-order QPT 
determined by solving the following equation, 
\begin{eqnarray}\label{cpinf}
h=4J\sinh(\eta)\sum_{n=-\infty}^\infty\frac{(-1)^n}{\cosh(n\eta)},
\end{eqnarray}
with $\eta = \cosh^{-1}(\Delta_{inf})$. The other CP, called $\Delta_1$, is
a first-order QPT given by
\begin{eqnarray}\label{cp1}
\Delta_1 = \frac{h}{4J}-1.
\end{eqnarray}

\begin{figure}%[!tb]
\centerline{\psfig{file=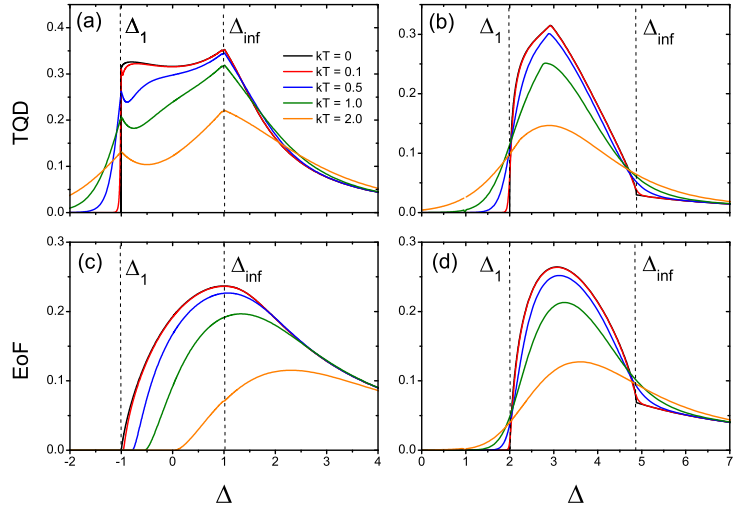,width=4.00in}}
\caption{TQD (panels a and b) and EoF (panels c and d) as functions 
of the tuning parameter $\Delta$ for $h=0$ (panels a and c) and $h=12$ 
(panels b and d). From top to bottom in the region where 
$\Delta_1<\Delta<\Delta_{inf}$, $kT=0,0.1,0.5,1.0,2.0$.}
\label{fig4}
\end{figure}

The behavior of TQD and EoF for the XXZ model at finite $T$ 
and $h=0$ was initially studied in ref.~\refcite{werPRL}. For $h=0$ 
the XXZ model has two CPs\cite{TAKAHASHI}. At $\Delta_{inf}=1$ 
the ground state changes from an XY-like phase ($-1<\Delta<1$) 
to an Ising-like antiferromagnetic phase ($\Delta>1$). 
At $\Delta_1=-1$ it changes from a ferromagnetic phase ($\Delta<-1$) to 
the critical antiferromagnetic phase ($-1<\Delta<1$). In ref.~\refcite{werPRL} 
we analyzed the behavior of TQD and EoF only for $\Delta>0$. 
Here we extend those results by computing the correlation functions 
for the remaining values of the anisotropy parameter, namely, $\Delta<0$. 
In Fig.~\ref{fig4} we plot 
TQD (panel a) and EoF (panel c) as a function of $\Delta$ for 
different values of $kT$ and $h=0$. 
For $T=0$ we can see that both 
TQD and EoF are able to detect the CPs. TQD is discontinuous 
at $\Delta_1$ while at $\Delta_{inf}$ the first-order derivative of 
TQD presents a discontinuity. Furthermore, EoF is zero for 
$\Delta<\Delta_1$ and non-zero for  $\Delta>\Delta_1$ reaching a 
maximum value at $\Delta=\Delta_{inf}$. However, as the temperature 
increases the maximum value of EoF is shifted to the right. Besides, 
EoF becomes zero also for $\Delta>\Delta_1$ as we increase $T$. 
On the other hand, the first-order derivative of TQD is still discontinuous at 
$\Delta_{inf}=1$ for finite $T$. We can also observe that TQD 
increases for $\Delta<-1$ as $T$ increases while 
its first-order derivative diverges at the CP $\Delta=-1$. As mentioned 
in Ref.~\refcite{werPRL} the cusp-like behavior at CPs $\Delta_1=1$ 
and $\Delta_{inf}=-1$ is due to an exchange in the set of projectors 
that minimizes the quantum conditional entropy (\ref{qce}).

The study about the XXZ model was further explored in 
Ref.~\refcite{werPRA}, with the addition of an external 
field. The effects of the magnetic field $h$ on the 
quantum correlations are exemplified in 
Fig.~\ref{fig4} (panels b and d), where we set $h=12$. 
The values of the CPs for $h=12$ are calculated employing 
Eqs. (\ref{cpinf}) and (\ref{cp1}), resulting in 
$\Delta_{inf}\approx4.88$ and $\Delta_1=2$. Again, 
for $T=0$ the CPs are detected by both TQD and EoF 
and, differently from the case $h=0$, the behavior of 
these two quantities is quite similar. Both quantities 
are zero for $\Delta<2$ and non-zero for $\Delta>2$, with 
their first-order derivatives diverging at the CP $\Delta_1=2$. 
However, the infinite-order QPT is no longer characterized by a 
global maximum of TQD or EoF, but by a discontinuity in their first-order 
derivatives. Note also that TQD presents a cusp-like behavior 
between the CPs. This behavior is once again related to the minimization 
procedure of the quantum conditional entropy and so far it is 
not associated to any known QPT for this model. 
It is important to mention that entanglement measures may 
also have a discontinuity and/or a divergence in their 
derivatives that are not related to a QPT\cite{yang}.

\begin{figure}%[!tb]
\centerline{\psfig{file=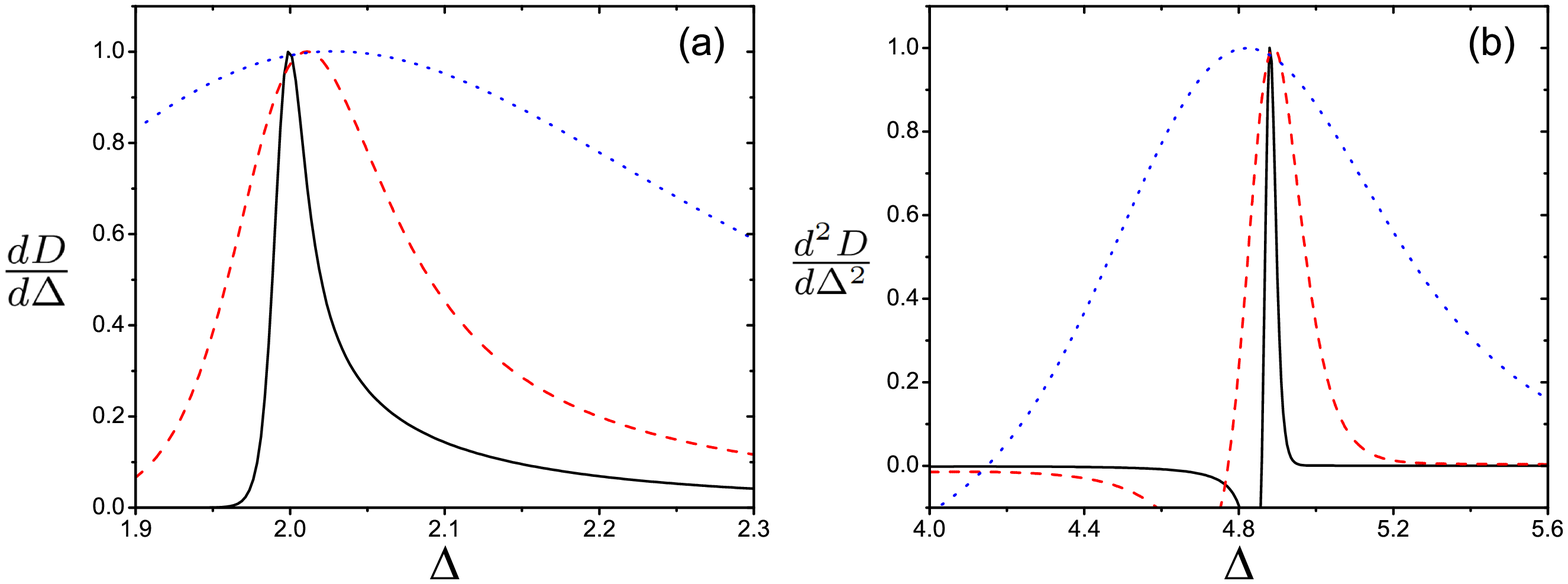,width=4.0in}}
\caption{First-order (panel a) and second-order (panel b) derivatives 
of TQD as functions of $\Delta$ for the XXZ model with $h=12$ and 
for $kT = 0.02$ (solid line), $kT=0.1$ (dashed line), and $kT = 0.5$ (dotted line). 
The derivatives plotted here are normalized, that is, each curve was 
divided by the maximum value of the respective derivative. The 
maximum of the first-order and second-order derivatives of TQD 
are very close to the CPs $\Delta_1=2$ 
and $\Delta_{inf}\approx4.88$, respectively.}
\label{fig5}
\end{figure}
Now we move on to the cases where $T>0$. When $T$ increases both curves 
of TQD and EoF become smoother and broader, with well defined derivatives 
in the CPs. Besides, the cusp-like behavior of TQD previously mentioned tends 
to disappear while both maximums of TQD and EoF decrease\cite{werPRA}.
We noted in Ref.~\refcite{werPRA} that some features of the derivatives 
of these quantities remain for a finite, but not too high temperature. 
To illustrate this fact, we plotted in Fig.~\ref{fig5} the first-order 
(panel a) and second-order (panel b) derivatives of TQD with respect 
to the anisotropy parameter $\Delta$ for $h=12$ and $kT=0.02, 0.1, 0.5$. 
To plot the curves for different temperatures in the same graph we 
normalized the derivatives of TQD, that is, for each $T$ we plotted 
the derivative of TQD divided by the maximum value of the respective 
derivative. For $T=0$ the divergence in the first-order derivative of both 
TQD and EoF spotlights the CP $\Delta_1$ while the CP $\Delta_{inf}$ is 
characterized by a divergence in the second-order derivative. 
As can be seen in Fig.~\ref{fig5}, although the divergence at the 
CPs disappears as $T$ increases, the derivatives reach their 
maximum values around the CPs. We used these maximum values to 
estimate the CPs at finite temperatures. The same analysis was 
applied to estimate the CPs using EoF instead of TQD. 
In Ref.~\refcite{werPRA} 
we compared the ability of TQD, EoF, and pairwise correlations 
($\med{\sigma_1^z\sigma_2^z}$ and $\med{\sigma_1^x\sigma_2^x}$) 
to point out the CPs for $T>0$ and we showed that TQD is the 
best candidate to estimate the CPs. 

\begin{figure}%[!tb]
\centerline{\psfig{file=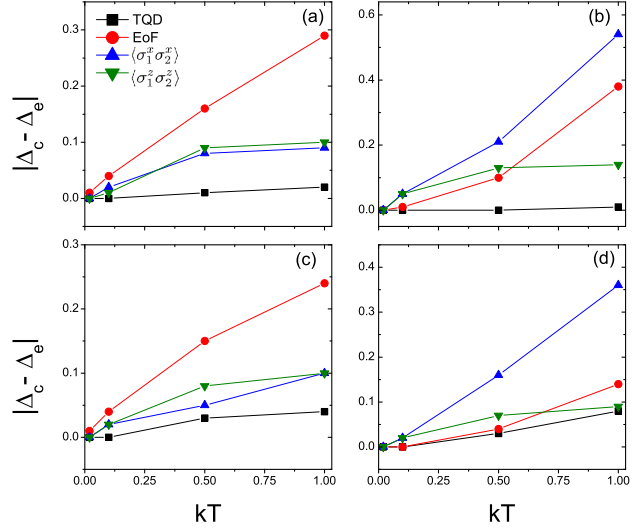,width=3.4in}}
\caption{The difference between the correct CP and the 
CP estimated by TQD (square), EoF (circle), $\med{\sigma_1^x\sigma_2^x}$ (up arrow), 
and $\med{\sigma_1^z\sigma_2^z}$ (down arrow) as a function of $kT$. 
In (a) and (b) we have $h=6$ with a first (a) and an infinite-order 
(b) CP; in (c) and (d) we have $h=12$ with a first (c) and an infinite-order 
(d) CP. $\Delta_c$ denotes the correct value of CP while $\Delta_e$ 
denotes the value of  CP estimated by the extremum values of the 
derivatives of the quantities involved.}
\label{fig6}
\end{figure}
To illustrate such result 
we compared in Fig.~\ref{fig6} the difference between the correct 
CP $\Delta_c$ and the CP estimated by our method $\Delta_e$ for 
$h=6$ and $h=12$. One can see in this figure that from zero to 
$kT\approx 1$ the CPs estimated by TQD are closer to the correct ones
than the estimated CPs coming from other quantities.

\subsection{XY Model}

The Hamiltonian of the one-dimensional XY model in a transverse field is given by 
\begin{eqnarray}
\label{HXY}
H_{xy}&=& - \frac{\lambda}{2} \sum_{j=1}^L 
\left[(1+\gamma)\sigma^x_j\sigma^x_{j+1}
+(1-\gamma)\sigma^y_j\sigma^y_{j+1}\right] 
-\sum_{j=1}^L\sigma^z_j,
\end{eqnarray}
where $\lambda$ is the strength of the inverse of 
the external transverse magnetic field and $\gamma$ 
is the anisotropy parameter. The transverse Ising model 
is obtained for $\gamma=\pm1$ while $\gamma=0$ corresponds 
to the XX model in a transverse field\cite{LSM}. 
At $\lambda_c=1$ the XY model undergoes a second-order 
QPT (Ising transition\cite{isingQPT}) that separates a 
ferromagnetic ordered phase from a quantum paramagnetic 
phase. Another second order QPT is observed for $\lambda>1$ 
at the CP $\gamma_c=0$ (anisotropy transition\cite{LSM,anisQPT}). 
This transition is driven by the 
anisotropy parameter $\gamma$ and separates a ferromagnet 
ordered along the $x$ direction and a ferromagnet ordered 
along the $y$ direction. These two transitions are of the same 
order but belong to different universality classes \cite{LSM,anisQPT}.

The XY Hamiltonian is $Z_2$-symmetric and can be exactly 
diagonalized \cite{LSM} in the thermodynamic limit $L\rightarrow\infty$. 
Due to translational invariance the two spin density operator $\rho_{i,j}$ 
for spins $i$ and $j$ at thermal equilibrium is \cite{osborne}
\begin{eqnarray}\label{doXY}
\rho_{0,k}&=&\frac{1}{4}\left[I_{0,k}
+\left\langle \sigma^z\right\rangle\left(\sigma^z_0+\sigma^z_k\right)\right]  
+ \frac{1}{4} \sum_{\alpha=x,y,z} 
\left\langle \sigma^\alpha_0\sigma^\alpha_k\right\rangle \sigma^\alpha_0\sigma^\alpha_k,
\end{eqnarray}
where $k=\left|j-i\right|$ and $I_{0,k}$ is the identity operator of dimension four.  
The transverse magnetization $\left\langle \sigma^z_k\right\rangle$
$=$ $\left\langle \sigma^z\right\rangle$ is
\begin{eqnarray}\label{tmag}
\left\langle \sigma^z\right\rangle=-\int_0^\pi 
(1+\lambda\cos{\phi})\tanh{(\beta\omega_\phi)}\frac{d\phi}{2\pi\omega_\phi},
\end{eqnarray}
with $\omega_\phi=\sqrt{(\gamma\lambda\sin{\phi})^2+(1+\lambda\cos{\phi})^2}/2$. 
The two-point correlation functions are given by
\begin{eqnarray}\label{tpcf}
\left\langle \sigma^x_0\sigma^x_k\right\rangle &=& \left|
\begin{array}{cccc}
G_{-1} & G_{-2} & \cdots & G_{-k}\\
G_{0} & G_{-1}  & \cdots & G_{-k+1} \\
\vdots & \vdots & \ddots & \vdots \\
G_{k-2} & G_{k-3} & \cdots & G_{-1}  \\
\end{array}
\right|,\\
\left\langle \sigma^y_0\sigma^y_k\right\rangle &=& \left|
\begin{array}{cccc}
G_{1} & G_{0} & \cdots & G_{-k+2}\\
G_{2} & G_{1}  & \cdots & G_{-k+3} \\
\vdots & \vdots & \ddots & \vdots \\
G_{k} & G_{k-1} & \cdots & G_{1}  \\
\end{array}
\right|,\\
\left\langle \sigma^z_0\sigma^z_k\right\rangle &=& 
\left\langle \sigma^z\right\rangle^2 - G_k G_{-k},
\end{eqnarray}
where
\begin{eqnarray*}
G_k&=&\int_0^\pi \tanh{(\beta\omega_\phi)}\cos{(k\phi)}(1+\lambda\cos{\phi}) 
\frac{d\phi}{2\pi\omega_\phi}\\
&-&\gamma\lambda\int_0^\pi \tanh{(\beta\omega_\phi)} \sin{(k\phi)\sin{\phi}} 
\frac{d\phi}{2\pi\omega_\phi}.
\end{eqnarray*}

The relation between TQD and QPT for 
the Ising model (XY model with $\gamma=1$) at $T=0$ was 
investigated initially by Dillenschneider\cite{Dil08} for 
first and second nearest-neighbors. More general results were 
obtained in Ref.~\refcite{Sar10} where TQD and EoF from 
first to fourth nearest-neighbors was computed for different 
values of $\gamma$. This study at $T=0$ showed that while EoF between 
far neighbors becomes zero, QD is not null and detects the QPT. The 
effects of the symmetry breaking process in entanglement and QD for the XY
and the XXZ models were discussed in Refs.~\refcite{Ami10}, where the 
low temperature regime was taken into account. In Ref.~\refcite{werPRA} 
we compared the ability of TQD and EoF for first and second nearest-neighbors 
to detect the CPs for the XY model at finite temperature.

\begin{figure}%[!tb]
\centerline{\psfig{file=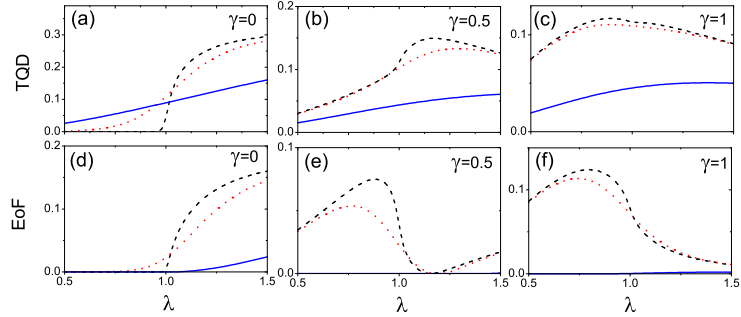,width=4.0in}}
\caption{(a)-(c) TQD and (d)-(f) EoF as functions of 
$\lambda$ for $kT=0.01$ (black/dashed line), $kT=0.1$ (red/dotted line) and 
$kT=0.5$ (blue/solid line) for nearest-neighbors. 
We use three values of $\gamma$ as shown in the graphs.}
\label{fig7}
\end{figure}

\begin{figure}%[!tb]
\centerline{\psfig{file=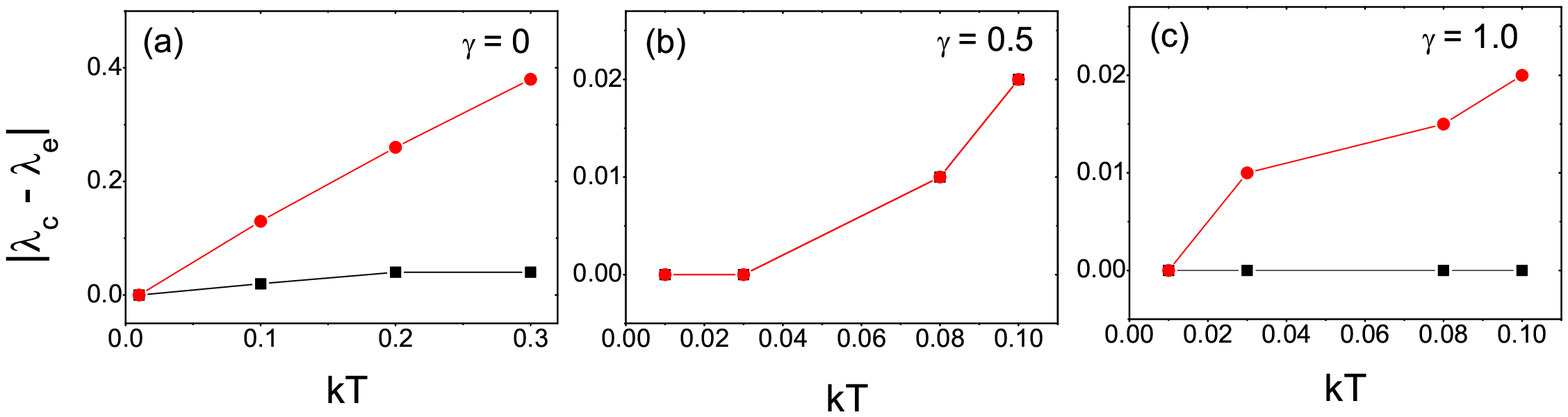,width=4.25in}}
\caption{The difference between the correct CP, $\lambda_c$, and, $\lambda_e$, 
the CP estimated either by TQD (square) or EoF (circle) as a function 
of $kT$ for (a) $\gamma=0$, (b) $\gamma=0.5$, and (c) $\gamma=1.0$.
Note that both curves coincide at panel (b).}
\label{fig8}
\end{figure}

The behavior of TQD and EoF for first nearest-neighbors as a 
function of $\lambda$ for $kT=0.01,0.1,0.5$ and $\gamma=0,0.5,1.0$ 
can be seen in Fig.~\ref{fig7}. First, note that TQD is more robust 
to temperature increase than EoF. For $kT=0.5$ TQD is 
always non-zero while EoF is zero or close to zero for almost all 
$\lambda$ (see the blue/solid curves in Fig.~\ref{fig7}). 
As showed in Ref.~\refcite{werPRA}, for second nearest-neighbors the 
situation is more drastic since EoF is always zero for $kT=0.5$. Now, 
to estimate the CPs at finite $T$ we used the same procedure adopted for 
the XXZ model. If the first-order derivative of TQD or EoF is divergent at 
$T=0$ then the CP is pointed out by a local maximum or minimum at $T>0$; 
if the first-order derivative is discontinuous at $T=0$ then we look after 
local maximum or minimum in the second derivative for $T>0$. These extreme 
values act as indicators of QPTs. The CPs estimated with such 
method are denoted by $\lambda_e$ while the correct CPs are denoted 
by $\lambda_c$. The differences between $\lambda_c$ and $\lambda_e$ as a 
function of kT for $\gamma=0,0.5,1.0$ are plotted in Fig.~\ref{fig8}. We can 
see in this figure that TQD provides a better estimate of the CP $\lambda_c=1$ 
than EoF. 

For $\gamma=0.5$ EoF and TQD give almost the same estimation of the CP
and for $\gamma=1.0$ TQD is better than EoF with predictions differing at
the second decimal place. For $\gamma = 0$, TQD outperforms 
EoF already in the first decimal place. Moreover, for $\gamma=0$ TQD 
is able to correctly estimate the CP for higher temperatures than 
for $\gamma=0.5$ and $\gamma=1.0$.

\begin{figure}%[!tb]
\centerline{\psfig{file=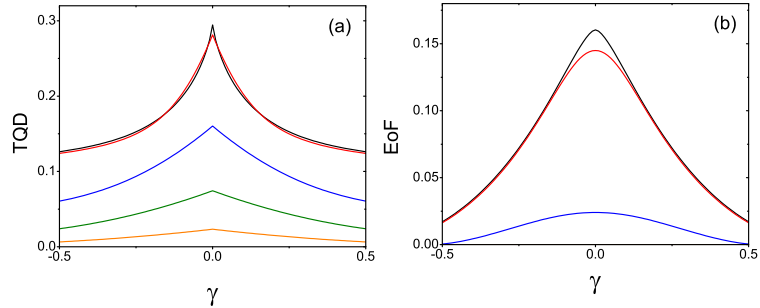,width=4.00in}}
\caption{(a) TQD and (b) EoF for the first nearest-neighbors as 
functions of $\gamma$. From top to bottom $kT=0.001, 0.1, 0.5, 1.0$, and
$2.0$. Here we fixed $\lambda = 1.5$ and the CP is $\gamma_c=0$.}
\label{fig9}
\end{figure}

So far we have studied a QPT driven by the magnetic field. However, 
for $\lambda>1$ the XY model undergoes a QPT driven by the 
anisotropy parameter $\gamma$, whose critical point is $\gamma_c=0$. 
To study such transition we fixed $\lambda=1.5$. In Fig.~\ref{fig9} 
we plotted TQD and EoF for the first-neighbors as 
functions of $\gamma$ and for $kT = 0.001, 0.1, 0.5, 1.0,$ and $2.0$. 
Note that the maximum of TQD and EoF is reached at the CP $\gamma_c=0$. 
However, only TQD has a cusp-like behavior at the CP. 
This 
pattern of TQD (maximum with a cusp-like behavior) remains up to $kT=2.0$. 
On the other hand, the maximum of EoF at the CP can only be 
seen as far as $kT<1$ for above this temperature EoF becomes zero. 
In Ref.~\refcite{werPRA} we computed TQD and EoF for second-neighbors. 
In this case TQD is able to detect the CP even for values 
near $kT=1.0$ while EoF is nonzero only for $kT\lesssim 0.1$.

\section{Conclusions}

In this article we presented a review of our studies about the 
behavior of quantum correlations in the context of spin chains at 
finite temperatures. The main goal of this paper was to analyze 
the quantum correlation's ability to pinpoint the critical points
associated to quantum phase transitions assuming  
the system's temperature is greater than the absolute zero. The two 
measures of quantum correlations studied here were 
quantum discord and entanglement, with the former producing the
best results. 

We first reviewed a work of two of us about a simple but illustrative
two spin-1/2 system described by 
the XYZ model with an external magnetic field\cite{Wer10}. We showed many surprising 
results about the thermal quantum discord's behavior. For example, and 
differently from entanglement, thermal quantum discord can increase with temperature in 
the absence of an external field even for such a small system; and that
there are situations where thermal quantum discord increases while
entanglement decreases with temperature. 
Furthermore, for the 
XXX model we observed for the first time that quantum discord could 
be a good candidate to signal a critical point at finite $T$. 

To check whether quantum discord is indeed a good critical point detector 
for temperatures higher than absolute zero, we analyzed its behavior
for an infinite chain described by the XXZ model 
without\cite{werPRL} and with\cite{werPRA} 
an external magnetic field and in equilibrium with a 
thermal reservoir at temperature $T$. Here we also extended our previous results by 
computing the quantum correlations between the first nearest-neighbor spins 
for the whole range of the anisotropy parameter (positive and negative values). 
In this way we were able to describe the two critical points of the XXZ model in the 
thermodynamic limit and the behavior of quantum discord near them. 
The results presented here and in refs.~\refcite{werPRL,werPRA} 
showed that quantum discord is far better the best critical 
point detector for $T>0$ with respect to all quantities tested (entanglement,
entropy, specific heat, magnetic susceptibility, and the two-point correlation
functions). 

Another model considered in this work was the XY model in a transverse 
magnetic field\cite{werPRA}. Again, we computed the quantum correlations between the first 
nearest-neighbors assuming the system in equilibrium with a thermal 
reservoir at temperature $T$. This model has two second-order quantum 
phase transitions, namely, an Ising transition and an anisotropy transition. 
For the Ising transition we observed that the critical point is better 
estimated by quantum discord. For the anisotropy transition both quantities,
entanglement and discord, provide an excellent estimate for the critical 
point at low temperatures. 
However, since for increasing temperatures quantum discord is more robust 
than entanglement, the former was able to 
spotlight the quantum critical point for a wider range of temperatures,
even for values of temperature where entanglement was absent. 

In conclusion, we showed that for the spin models studied here and in 
Refs.~\refcite{werPRL,werPRA}
quantum discord was the best quantum critical point estimator among all 
quantities tested when the system assumes a finite temperature. 
It is also important to mention that the knowledge of the order 
parameter was not needed to estimate the critical points. Therefore, 
we strongly believe that our results suggest that 
quantum correlations - mainly quantum discord -  are important 
tools to study quantum phase transitions in realistic scenarios, 
where the temperature is always above absolute zero.

\section*{Acknowledgements}

TW and GR thank the Brazilian agency CNPq (National Council for Scientific and Technological 
Development) for funding and GAPR thanks CNPq and FAPESP (State of S\~ao Paulo Research 
Foundation) for funding. GR also thanks CNPq/FAPESP 
for financial support through the National Institute of Science and Technology 
for Quantum Information.


\begin{thebibliography}{0}

\bibitem{sac99} S. Sachdev, {\it Quantum Phase Transitions} (Cambridge University Press, Cambridge, 1999).

\bibitem{row10} S. Rowley \textit{et al.}, Phys. Status Solid B \textbf{247}, 469 (2010).

\bibitem{dol10} V. F. Gantmakher and V. T. Dolgopolov, Physics-Uspekhi \textbf{53}, 1 (2010).

\bibitem{gre02} M. Greiner \textit{et al.}, Nature (London) \textbf{415}, 39 (2002).

\bibitem{pres00} J. Preskill, J. Mod. Opt. \textbf{47}, 127 (2000).

\bibitem{Nie99} T. J. Osborne and M. A. Nielsen, Phys. Rev. A \textbf{66}, 032110
(2002); A. Osterloh \textit{et al.}, Nature(London) \textbf{416}, 608 (2002);
J. I. Latorre, E. Rico, and G. Vidal, Quantum Inf. Comp. \textbf{4},
48 (2004); G. Vidal \textit{et al.}, Phys. Rev. Lett. \textbf{90}, 227902 (2003);
R. Somma \textit{et al.}, Phys. Rev. A \textbf{70}, 042311 (2004);
T. -C. Wei \textit{et al.}, Phys. Rev. A \textbf{71}, 060305 (2005);
F. Verstraete, M. Popp, and J. I. Cirac, Phys. Rev. Lett. {\bf 92}, 027901 (2004); 
M. Popp \textit{et al.}, Phys.Rev. A {\bf 71}, 042306 (2005).

\bibitem{lidar} L.-A. Wu, M. S. Sarandy, and D. A. Lidar, Phys. Rev. Lett. \textbf{93}, 250404 (2004).

\bibitem{oliveira} T. R. de Oliveira et al., Phys. Rev. Lett. \textbf{97}, 170401 (2006); 
T. R. de Oliveira, G. Rigolin, and M. C. de Oliveira, Phys. Rev. A \textbf{73}, 010305(R) (2006); 
G. Rigolin, T. R. de Oliveira, and M. C. de Oliveira, Phys. Rev. A \textbf{74}, 022314 (2006).

\bibitem{zurek} H. Ollivier and W. H. Zurek, Phys. Rev. Lett. \textbf{88}, 017901 (2001).

\bibitem{vedral} L. Henderson and V. Vedral, J. Phys. A: Math. Gen. \textbf{34}, 6899 (2001).

\bibitem{Dil08} R. Dillenschneider, Phys. Rev. B \textbf{78}, 224413 (2008).

\bibitem{disT0} M. S. Sarandy, Phys. Rev. A \textbf{80}, 022108 (2009); 
C. C. Rulli and M. S. Sarandy, Phys. Rev. A \textbf{81}, 032334 (2010); 
C. C. Rulli and M. S. Sarandy, Phys. Rev. A \textbf{84}, 042109 (2011); 
A. Saguia et al. , Phys. Rev. A \textbf{84}, 042123 (2011); 
M. Allegra, P. Giorda, and A. Montorsi, Phys. Rev. B \textbf{84},  245133 (2011).

\bibitem{Ved01} M. C. Arnesen, S. Bose, and V. Vedral, Phys. Rev. Lett. \textbf{87}, 017901 (2001).

\bibitem{Kam02} G. L. Kamta and A. F. Starace, Phys. Rev. Lett. \textbf{88}, 107901 (2002);
X. Wang, Phys. Rev. A \textbf{64}, 012313 (2001); 
Y. Sun, Y. Chen, and H. Chen, Phys. Rev. A \textbf{68}, 044301 (2003);
S-J. Gu, H-Q. Lin, and Y-Q. Li, Phys. Rev. A \textbf{68}, 042330 (2003).

\bibitem{Rig03} G. Rigolin, Int. J. Quant. Inf. \textbf{2}, 393 (2004).

\bibitem{ltte} L. Amico and D. Patan\'e, Europhys. Lett. \textbf{77}, 17001 (2007).

\bibitem{reviews} L. Amico \textit{et al.}, Rev. Mod. Phys. \textbf{80}, 517 (2008); 
L. Amico and R. Fazio, J. Phys. A: Math. Theor. \textbf{42}, 504001 (2009). 

\bibitem{Wer10} T. Werlang and G. Rigolin, Phys. Rev. A \textbf{81}, 044101 (2010).      

\bibitem{werPRL} T. Werlang, C.Trippe, G.A.P. Ribeiro and G. Rigolin, Phys. Rev. Lett. \textbf{105}, 095702 (2010).                          

\bibitem{werPRA} T. Werlang, G.A.P. Ribeiro and G. Rigolin, Phys. Rev. A \textbf{83}, 062334 (2011).  

\bibitem{EPR}  A. Einstein, B. Podolsky, and N. Rosen, Phys. Rev. \textbf{47}, 777 (1935).

\bibitem{Bel64} J. S. Bell, Physics \textbf{1}, 195 (1964).

\bibitem{Wer89} R. F. Werner, Phys. Rev. A \textbf{40}, 4277 (1989).

\bibitem{nielsen} M. A. Nielsen and I. L. Chuang, 
{\it Quantum Computation and Quantum Information}. (Cambridge University Press, Cambridge, 2000).

\bibitem{Hor09} R. Horodecki \textit{et al.}, Rev. Mod. Phys. \textbf{81}, 865 (2009).

\bibitem{ved10b} K. Modi \textit{et al.}, Phys. Rev. Lett. \textbf{104}, 080501 (2010).

\bibitem{brule} A. Peres, {\it Quantum Theory: Concepts and Methods}. (Kluer Academic Publishers, New York, 2002). 

\bibitem{ved10} B. Dakic, V. Vedral, C. Brukner, Phys. Rev. Lett. \textbf{105}, 190502 (2010).

\bibitem{minDIS} S. Hamieh, R. Kobes, and H. Zaraket, Phys. Rev. A \textbf{70}, 052325 (2004).

\bibitem{Dnum} D. Girolami and G. Adesso, Phys. Rev. A \textbf{83}, 052108 (2011).

\bibitem{Danalitico} M. Ali, A. R. P. Rau, and G. Alber, Phys. Rev. A \textbf{81}, 042105 (2010); 
ibid. \textbf{82}, 069902(E) (2010).

\bibitem{Woo98} W. K. Wootters, Phys. Rev. Lett. \textbf{80}, 2245 (1998).

\bibitem{Ron12} X. Rong \textit{et al.}, eprint arxiv:1203.3960v1 [quant-ph].

\bibitem{NLIE} M. Bortz and F. G\"ohmann, Eur. Phys. J. B \textbf{46}, 399 (2005); 
H.E. Boos \textit{et al.}, J. Stat. Mech. (2008) P08010; 
C. Trippe, F. G\"ohmann, and A. Kl\"umper, Eur. Phys. J. B \textbf{73}, 253 (2010).

\bibitem{GAUDIN} J. Cloizeaux and M. Gaudin, J. Math. Phys. \textbf{7}, 1384 (1966).

\bibitem{TAKAHASHI} M. Takahashi, {\it Thermodynamics of one-dimensional solvable models} 
(Cambridge University Press, Cambridge, England, 1999).

\bibitem{yang} Min-Fong Yang, Phys. Rev. A \textbf{71}, 030302 (2005).

\bibitem{LSM} E. Lieb, T. Schultz, and D. Mattis, Ann. Phys. \textbf{16}, 407 (1961); 
E. Barouch, B.M. McCoy, and M. Dresden, Phys. Rev. A \textbf{2}, 1075 (1970); 
E. Barouch and B.M. McCoy, Phys. Rev. A \textbf{3}, 786 (1971).

\bibitem{isingQPT} P. Pfeuty, Ann. Phys. (New York) \textbf{57}, 79 (1970).

\bibitem{anisQPT} M. Zhong and P. Tong, J. Phys. A: Math. Theor. \textbf{43}, 505302 (2010). 

\bibitem{osborne} T. J. Osborne and M. A. Nielsen, Phys. Rev. A \textbf{66}, 032110 (2002).

\bibitem{Sar10} J. Maziero \textit{et al.}, Phys. Rev. A \textbf{82}, 012106 (2010).

\bibitem{Ami10} O. F. Sylju\aa sen, Phys. Rev. A \textbf{68}, 060301(R) (2003);
O. F. Sylju\aa sen, Phys. Lett. A \textbf{322}, 25 (2004);
A. Osterloh, G. Palacios, and S. Montangero, Phys. Rev. Lett. \textbf{97}, 257201 (2006);  
T. R. de Oliveira et al., Phys. Rev. A \textbf{77}, 032325 (2008);
B. Tomasello \textit{et al}, Europhys. Lett. \textbf{96}, 27002 (2011).


\end{thebibliography}
\end{document}